# The Alignment of High-resolution Solar Prominence Images Observed by the New Vacuum Solar Telescope

Yunfang Cai[1,2], Yongyuan Xiang[1,2], and Kaifan Ji[1]
[1] Yunnan Observatories, Chinese Academy of Sciences, Kunming Yunnan 650216, People's Republic of China
[2] Yunnan Key Laboratory of Solar Physics and Space Science, Kunming Yunnan 650216, People's Republic of China

## Abstract

High spatial resolution observation of solar prominence is an important observation subject of the New Vacuum Solar Telescope (NVST). While the current level of observation and image reconstruction technologies for solar prominences are advanced, a significant challenge remains in achieving high-precision alignment among high-resolution prominence images observed at different times and different off-bands. Existing alignment approaches either become ineffective or yield low accuracy, and always require manual intervention during the alignment. These limitations are largely due to the stronger edge gradient and lower structural contrast of the prominence images compared with the solar surface ones. In response to this challenge, our study aims to develop an effective and robust algorithm for high-precision alignment of the prominence images. We thoroughly consider the unique structural characteristics of prominence images and the specific application conditions of various alignment algorithms. Consequently, we propose a comprehensive alignment method that incorporates the optical flow field of the solar surface, the gradient of the solar edge, and the cross-correlation within the solar prominence region. This method is designed to accurately determine the movement displacements among the prominence images. Our results demonstrate that this alignment method excels in both accuracy and robustness, making it well-suited for handling the diverse postures of solar prominence images captured by the NVST.

*Unified Astronomy Thesaurus concepts:* Astronomy data reduction (1861); Solar chromosphere (1479); Solar prominences (1519)

## 1. Introduction

Solar prominences are the special activity phenomena of the solar atmosphere observed on the solar limb, composed of low-temperature high-density plasma, suspended in extremely high temperature and thin corona (E. Tandberg-Hanssen 1995). Prominences typically appear as dark ribbon-like features and also are well-known as filaments when observed on the solar disk. Scientists usually study how these high-density materials can exist stably in hot and low-density corona using high-resolution imaging observations of prominences/filaments, which is one of the hot topics in solar physics (X. L. Yan et al. 2014). Since the high-resolution observation can provide an ultrafine view of the internal structures of prominences/filaments, consecutive images serve to improve our knowledge about their evolution and physical properties. The high-resolution prominence observations are favored by many solar physicists. However, before these high-resolution prominence data sets can be utilized for scientific research, they need to be carefully aligned. The high-resolution images are captured by the larger-aperture ground-based telescopes with a tunable narrow-band filtergram, such facilities include the Swedish Solar Telescope, Goode Solar Telescope (GST), New Vacuum Solar Telescope (NVST), and so on. They are capable of recording monochromatic images at different off-bands across a spectral line over successive wavelengths within a specific time interval. However, the high-resolution imaging observations are inherently subject to the influence of near-ground telluric atmospheric turbulence, which triggers random aberrations in the time series of images captured by these telescopes. Moreover, the heavy optical support system of telescopes can cause the guiding and tracking system to deviate slightly from the preset value during the operation. This results in random displacements in the captured images, affecting both the monochromatic sequence images at the same wavelength and the imaging spectroscopic images with different center wavelengths. To compensate for these distortions, adaptive optics (AO) systems (A. Wirth & R. Ruquist 1984; C. Rao et al. 2016) have been widely implemented in the larger-aperture ground-based telescopes. AO systems rely on the strong contrast structures of the solar photosphere layer, such as the TiO-band or continuum intensity images, to correct for aberrations in real time. However, these structures are not visible in prominence observations of the photosphere, rendering the AO system ineffective for these specific observations. Therefore, to achieve science-ready high-resolution prominence images, postimage reconstruction technology (G. Weigelt & B. Wirnitzer 1983; G. P. Weigelt 1977) is essential to overcome the random image aberrations induced by telluric atmospheric turbulence. Additionally, an alignment method is also necessary to correct the image displacements caused by the mechanical vibration of telescopes. The high-resolution prominence observations of NVST have achieved many excellent results in the field of solar physics research (Y. Shen et al. 2015; X. Yan et al. 2015; Y. Bi et al. 2020; B. Yang et al. 2021; J. Wang et al. 2022) due to its advanced observation conditions and reconstruction technology. Y.-Y. Xiang et al. (2016) and his team developed an improved algorithm in 2016 that utilizes speckle imaging technology to achieve the high-resolution reconstruction of prominence images. This method has been successful in reconstructing images with a spatial resolution of approximately 0″.3, which is close to the diffraction limit of the NVST to clearly display the internal ultrafine structure of solar prominences. However, the alignment of the high-resolution prominence images has always been a tough issue. Common alignment algorithms often struggle to achieve the necessary accuracy, and the alignment process is not only time-consuming but also requires manual intervention owing to the special structures of prominence images.

The observed solar images always have weak structure contrast, and strong noise, and contain many different active phenomena with random and nonrigid motion/evolution compared with the general natural images. The alignment of solar images is inherently a challenging problem and has been continuously researched. The cross-correlation algorithm, such as the global Fourier correlation tracker (GFCT), based on the regional similar feature structures between the reference and the moving image, is a prevalent alignment method for time-series solar images at the same wavelength (S. Berkebile-Stoiser et al. 2009; O. Kuehner et al. 2010). This approach is effective for the images with large active areas or slow-moving/evolving structures (e.g., solar sunspots, quiet filament, and so on) and can achieve subpixel alignment accuracy. For the images with extensive quiet regions or significant changes, the most direct solution is to find similar features (such as the distinctive points, lines, and regions) between the two images as the references to deduce transform parameters, which method is the so-called feature point matching (S. L. Guglielmino et al. 2010). The approach can simultaneously derive the transitional, scaling, and rotational transform parameters between the pair of images, which is particularly useful for solar images with different resolutions, even those acquired by different telescopes. However, its alignment accuracy completely depends on the number of feature points (P. L. Smith et al. 2006). In the past, manual alignment by means of artificially selecting feature points was commonly used; this process was usually time-consuming, and difficult to achieve high alignment accuracy due to insufficient feature points. To automate the extraction of feature points, algorithms like the scale-invariant feature transform (SIFT) method (D. Lowe 2004) and optical flow (OF) approach (B. K. P. Horn & B. G. Schunck 1981) have been widely used in solar image processing. For instance, K. Ji et al. (2019) applied the SIFT method to register the photospheric and chromospheric heliograms of NVST with the full-surface data of Helioseismic and Magnetic Imager and Global Oscillation Network Group; we proposed an OF-based algorithm to align winged Hα data collected by NVST (Y.-F. Cai et al. 2022) and evaluated its alignment accuracy with the raster images of Fast Imaging Solar Spectrograph operating at GST (W. Cao et al. 2010), and H. Liu et al. (2022) applied this method for high-resolution solar image reconstruction. X. Yang et al. (2022) also utilized these two methods to register and align the high-resolution imaging data acquired by GST. All those applications have achieved excellent results and have been successfully applied to conventional data observed by NVST and GST. However, these current approaches often fail when applied to align the high-resolution prominence images.

The high-resolution prominence image of NVST is local imaging of the Sun, which includes two main areas of solar surface and prominence. Figure 1 shows the high-resolution prominence images with various postures at the Hα line center observed by NVST, which have significant characteristics compared with the solar surface images and bring greater challenges for image alignment. These challenges are mainly discussed from the following four aspects: (1) The prominence images present a sudden decrease in intensity from the solar surface to the limb, leading to a strong gradient. When using the classic cross-correlation method to align the images, this gradient results in a very insignificant maximum value of the image correlation surface. Additionally, there is a noticeable autocorrelation between a pair of images along the solar edge, causing the aligned images to drift in the direction of the solar edge. (2) In a prominence image, the prominence's intensity is significantly weaker than the solar surface. To clearly display the internal morphology of prominence, the exposure time for prominence observation is usually increased. This results in the solar disk area being in a semisaturated state with reduced resolution and contrast, as shown in images (2) and (3) of Figure 1. Also, the solar surface region always occupies a very small area of the entire prominence image, making it difficult to extract enough feature points for high-precision image alignment using the feature point matching method. (3) The routine observations by the NVST often have to pause image acquisition due to various inevitable factors, such as cloudy weather, the limited position of the derotator, data storage, and so on. In order to coordinate the position of the derotator, observers place the target of interest (sunspots, filaments, prominences, etc.) at the center of the field of view (FOV), but cannot guarantee that the target's position and posture will be consistent across each group of data. Therefore, we often see that there are multiple groups of data with the same target on one day from the official website of NVST, and the posture of the target in different groups is random, as shown in the three images in the lower panel of Figure 1. This phenomenon results that images from different groups cannot be processed in batches and often require manual selection of appropriate regions for alignment, which is time-consuming and inconvenient for scientists utilizing the data set. (4) The NVST's prominence observations often include multiwavelength imaging spectroscopic images. Besides the images of the Hα line center, the off-band images are also collected within a short time. The intensity of the prominence region is weaker in off-band images than in line center ones, especially in the far winged. Figure 2 shows the high-resolution imaging spectroscopic prominence images of three wavelengths from NVST. Aligning the prominence images of multiple wavelengths is also an enormous challenge due to these intensity differences.

After consulting with expert data scientists and soliciting demands and suggestions from NVST data users, we urgently need to develop an effective algorithm, which should employ an automatic and highly accurate image alignment method to well-align the NVST's high-resolution prominence images. The goal is to generate a scientifically convenient data set for prominence studies. Our task involves aligning two types of high-resolution prominence images: first, monochromatic images captured at the same wavelength over a series of time intervals; and second, imaging spectroscopic images that span various wavelengths. Upon close examination, we have identified several stable characteristics within the frames of prominence images. These include certain structures on the solar surface, the edge of the Sun, and the shape of the prominence. Thereby, we can segment a prominence image into those three regions, and make full use of their features combined with different methods to well-align prominence images. The paper is organized as follows: In Section 2, the instrumentation and observations are described. The alignment methodologies and their corresponding detailed processing steps are exhibited in Section 3. In Section 4, we evaluate the alignment accuracy of our algorithm and present the alignment results. Finally, the conclusion and discussion are given in Section 5.

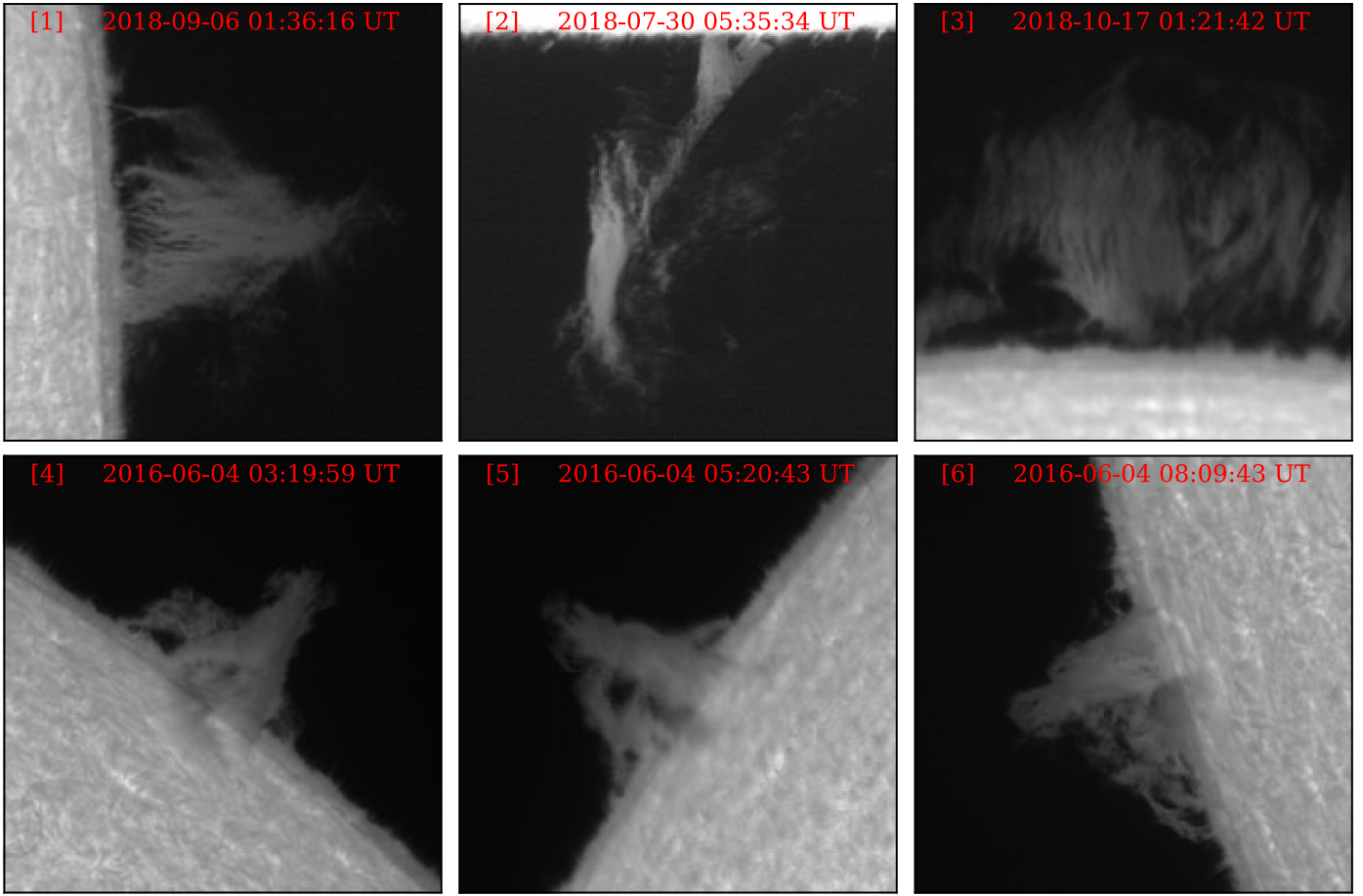

**Figure 1.** The high-resolution prominence images at the Hα line center with the different postures observed by NVST. The lower panel images come from different groups of data with the same target on one day.

## 2. Instrumentation and Observations

The NVST is a ground-based vacuum solar telescope equipped with a 985 mm clear aperture primary mirror, located at Fuxian Lake in Yunnan Province, China, administered by Fuxian Solar Observatory/Yunnan Observatories (Z. Liu et al. 2014). It can achieve scientific observation of the Sun using a multiband high-resolution imaging system (MHRIS; Y.-Y. Xiang et al. 2016) and a grating spectrometer system (Y. Cai et al. 2017), covering a wavelength ranges from 0.3 to 2.5 μm. The prominence images utilized in this study are captured by the Hα channel of the MHRIS, which employs a 2D imaging spectrometer based on a tunable Lyot filter with a bandwidth of 0.25 Å. The images are recorded using a PCO 2000 camera, which features an original sensor resolution of 2K ×2K pixels$^2$. For routine observations, a binning factor of 2 × 2 is applied, resulting in observation data with dimensions of 1024 × 1024 pixels$^2$. With an image scale of 0.″164 pixel$^{-1}$, each image covers an FOV of 168″ × 168″.

The raw prominence data from the NVST are first processed with conventional strength calibration techniques, including dark-current subtracted and flat-field corrected. To further mitigate the effects of atmospheric seeing, the data are reconstructed by the speckle masking technique (Z. Liu et al. 1998; Y.-Y. Xiang et al. 2016) to obtain the high-resolution monochromatic images of each off-band. A single high-resolution image is reconstructed from a series of 100 short exposure frames, with each frame having an exposure time of 10 or 20 ms. The interval between successive high-resolution images is contingent upon the number of off-bands selected for observation. Generally, a higher number of off-bands results in a longer interval time. The most off-band observations are the 3–5 bands mode, it always requires approximately 20–50 s to complete a single wavelength scanning. To demonstrate our alignment technique, we present two sets of high-resolution prominence data. The first set, observed on 2016 June 4, spans a relatively long duration from 03:00 to 08:30 UT. This set comprises three groups of prominence data, each exhibiting different postures, as shown in the lower panel images of Figure 1. Since only the Hα line center images were collected in this set, it serves to illustrate the alignment method for monochromatic sequence prominence images. The second set consists of imaging spectroscopic prominence images observed on 2016 August 17. This set includes a total of five center wavelengths: two Hα blue wings (−0.4 Å, and −0.7 Å), the Hα line center, and two Hα red wings (+0.4 Å, and +0.7 Å).

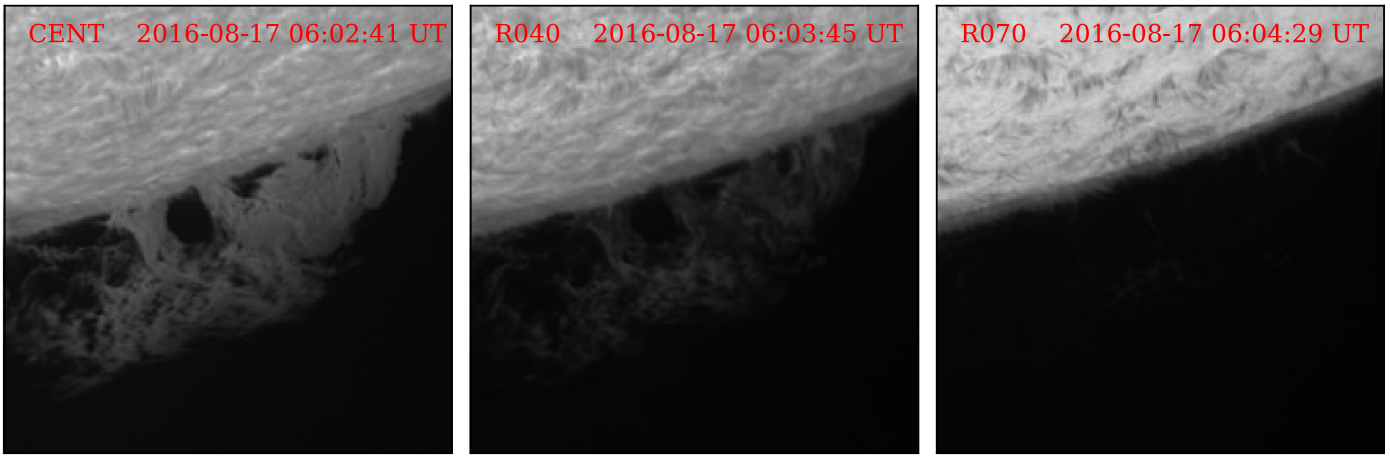

**Figure 2.** The high-resolution imaging spectroscopic prominence images of three wavelengths (CENT corresponds to the Hα line center, R040 to the red wing 0.4 Å, and R070 to the red wing 0.7 Å).

This data set is utilized to investigate the alignment approach for imaging spectroscopic prominence images.

### 3. Alignment Method for the Prominence Images

Given the current state of high-resolution prominence observations from the NVST, we aim to achieve high-precision alignment of these images by segmenting each image into three distinct regions: the solar surface area, the solar edge, and the solar prominence. As for the solar surface areas, they are usually relatively quiet regions with some small pores. The alignment algorithm of feature point matching based on the OF field has proven to be an effective method for these areas, offering the highest precision and the smallest cumulative error as detailed in the studies by Y.-F. Cai et al. ([2022](#)) and X. Yang et al. ([2022](#)). By employing the OF method, we can extract stable feature points from the solar surface areas to determine the image's movement displacement. The solar edge, which refers to the boundary of the solar photosphere excluding solar spicules, is a very stable characteristic in high-resolution prominence images. We utilize the intensity gradient of the solar edge to pinpoint the position of each image, providing a reliable reference for alignment. The prominence regions always have distinct outlines and abundant internal structures. The cross-correlation algorithm, based on a similar structure of prominence, is better suited for tracking these structures' translations. However, in our high-resolution prominence data, the solar surface area often has a lower resolution and occupies a small portion of the image; the solar edge is only a narrow part of the Sun's perimeter, and the solar prominence region has weak intensity and strong noise. Relying solely on any one of these regions and methods would be insufficient for achieving high alignment precision. Therefore, we propose to make full use of all three regions and combine them with various solutions to well-align the NVST's high-resolution prominence images.

How to accurately divide a prominence image into the aforementioned three regions is a significant challenge, as the target's posture is arbitrary, and the solar edge in an image is often inclined, as depicted in Figure [1](#). To facilitate batch processing of data from different groups and ensure the generality of our approach, we first preprocess the prominence images to orient them in a posture, namely rotating the edge of prominence to the "horizontal" direction. This standardized posture not only aids in the division of the prominence image into three parts but also benefits solar scientists by providing consistency in data products across different groups. To ensure precise alignment of the images of monochromatic sequences prominence and imaging spectroscopy from the NVST, we develop a comprehensive strategy. This strategy involves an initial preprocessing phase for the data, followed by the application of different processing methods with the three distinct parts. Figure [3](#) is a flowchart describing our alignment process and the main methods used for the solar prominence data. A detailed explanation of each method and its application to the data will be provided in the subsequent sections.

#### 3.1. Preprocessing for the Prominence Images

To orient images of solar prominence edges in a convenient posture, a straightforward approach is to define the solar center as a fixed reference point for image rotation. However, due to the limited FOV of our high-resolution prominence image, which covers only one-third of the Sun's surface, it is not feasible to capture both the solar center and the prominence within the same frame. To address this challenge, we achieve a "horizontal" orientation for the prominence edge in two steps: first, by extracting the coordinates of the solar edge from the prominence image, and second, by calculating the rotation angle of the solar edge relative to the horizontal axis. To standardize the orientations of prominence images across different data sets, we have established a protocol where, after preprocessing, the solar edge is aligned to a "horizontal" position, and the solar prominence is oriented in a downward direction within an image frame.

##### 3.1.1. Extract the Coordinations of Solar Edge

We illustrate the process of extracting the coordinates of the solar edge from a prominence image in Figure [4](#). The original image is denoted as M. We employ a morphological method to obtain the solar edge: First, the image M is filtered using the minimum threshold value, which effectively removes the prominence region and retains only the solar surface information, resulting in image M1 as shown in Figure [4](#). Next, the image M1 is then binarized and subjected to a connected-component analysis to eliminate scattered points, and morphological opening and closing operations are applied in succession, yielding a binary image M2 that includes the solar surface and edge. Then, the mage M2 undergoes morphological erosion with a 3 × 3 pixels$^2$ structuring element to remove the solar edge. The position coordinates (x, y) of the solar edge are determined by subtracting the eroded image from M2, and these coordinates are plotted as a red line in the M2 image. Given that the actual solar edge is smooth, we apply the parabola fitting with the random sample consensus (RANSAC) method for the position coordinates of the solar edge. The smoothed position coordinates (X, Y) represent the effective coordinates of the solar edge, which are depicted as a blue line in the M2 image.

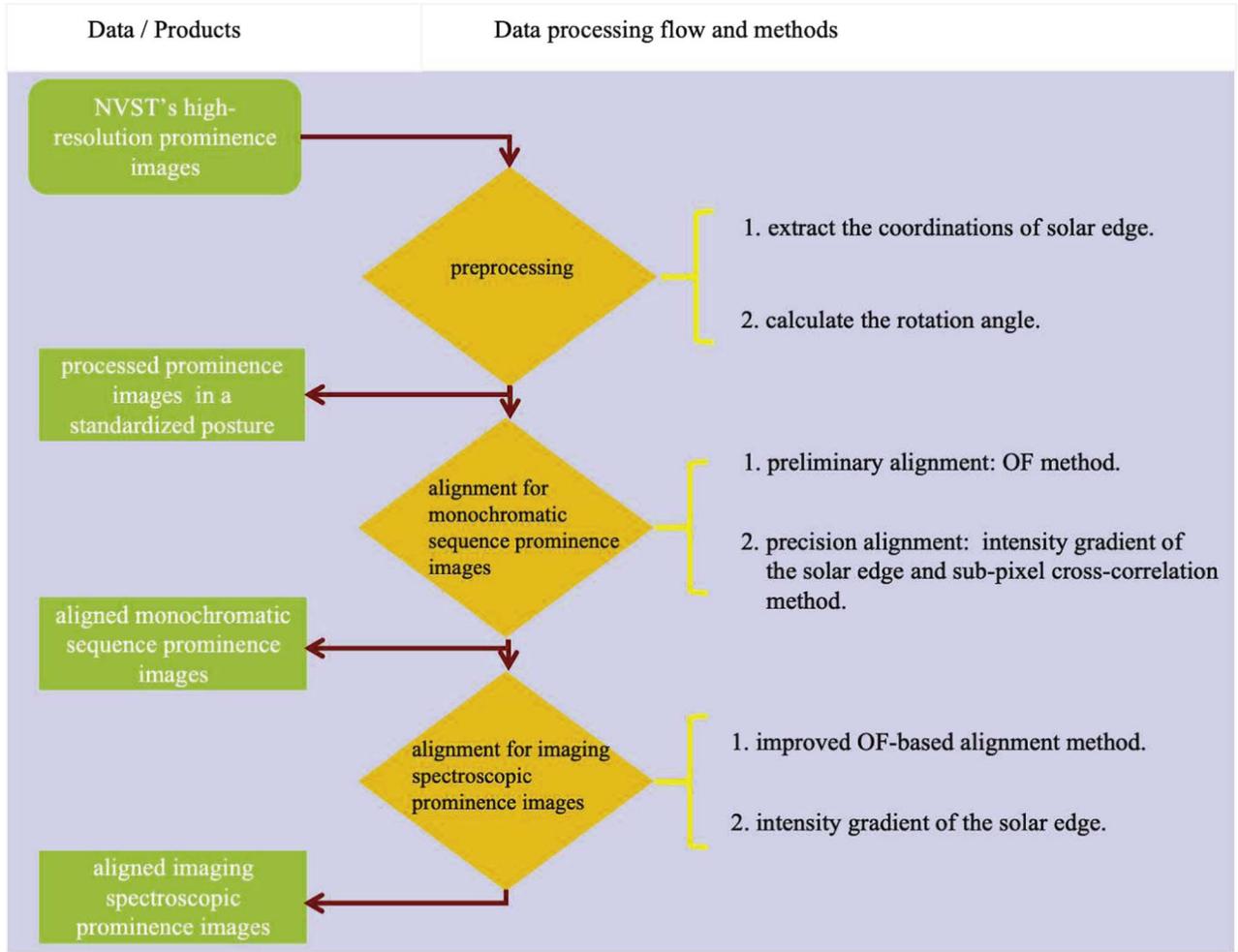

**Figure 3.** The NVST high-resolution prominence image alignment flowchart.

*3.1.2. Calculate the Rotation Angle of the Prominence Image*

The posture of the solar edge can be calculated using the Pearson correlation coefficient between the coordinates y and their RANSAN fitting values Y. Here, we employ the coarse-to-fine approach to rotate the solar edge to the "horizontal" direction accurately. The coarse step is to linear fit on the coordinates (X, Y) to obtain the approximate angle $\theta 1$ of the solar edge relative to the horizontal axis. The original image M from Figure 4 is then rotated by the angle $\theta 1$ to produce image M′ as shown in Figure 5. The fine step is to use the solar center to rotate the solar edge of M′ to the true "horizontal" direction. As well known, the radius of the Sun is 16′, and the image scale of NVST high-resolution prominence images is 0.″164 pixel$^{-1}$. When the solar radius $R_{Sun}$ is projected onto our images, it corresponds to approximately 5853.6 pixels. The edge coordinates (X′, Y′) of M′ can also be obtained with the same method as the above section. We then apply the radius constraint least-squares circle fitting method using the solar radius $R_{Sun}$ and the edge coordinates (X′, Y′) to determine the position coordinates of the solar center ($X_c$, $Y_c$), as well as the true edge coordinates ($X_t$, $Y_t$) using the equation of $R^2 = (x–X_c)^2+(y–Y_c)^2$. In addition, the coordinates of solar edges at the left and right sides of M′ are $X_{t\_min}$ and $X_{t\_max}$, respectively, and thus the rotation angle of $\theta 2$ around the solar center is calculated using the following equation:

$$\sin(\theta 2) = (((X_{t\_max} - X_{t\_min})/2 + X_{t\_min}) - X_c)/R. \quad (1)$$

The M′ is then rotated by the angle $\theta 2$ to produce image M″ as shown in Figure 5, where the solar edge of prominence images is "horizontal". Thus, from the initial image M to the M″, the total rotation angle of the prominence image is $\theta = \theta 1 + \theta 2$.

*3.2. Alignment of Monochromatic Sequence Prominence Images*

The well-aligned monochromatic sequence prominence images are essential for investigating the evolution of internal structures and activity phenomena within solar prominences.

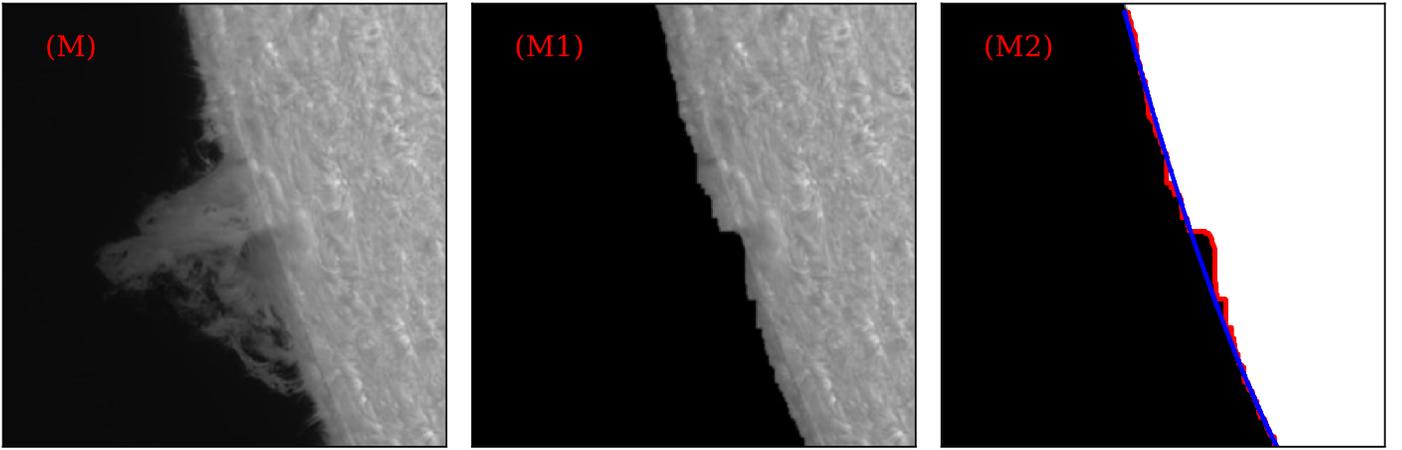

**Figure 4.** Extracting the coordinations of solar edge from a high-resolution prominence image.

After preprocessing, these images are rotated into a convenient posture, allowing them to be easily decomposed into three distinct regions: the solar surface, the solar edge, and the prominence areas. We perform the following methods with these three regions to align the monochromatic sequence prominence images. The first is preliminary alignment with the solar surface region using the OF method. We obtain the relative motion information of images by statistically analyzing the OF field within the solar surface region. This approach is chosen over applying the OF method to the entire image to avoid the influence of the OF fields from solar prominence and its surrounding blank areas on our statistical calculations of displacement, which could compromise the alignment accuracy. The second is precision alignment with the solar edge and the solar prominence area. The preliminary alignment accuracy is generally limited due to fewer feature points extracted from the solar surface area. The sequential images after preliminary alignment often show systematic drift along the solar edge direction, accompanied by random oscillation. To address this, we calculate the intensity gradient of the solar edge to determine its positions and further correct the preliminary displacement in the vertical direction. However, the horizontal variation of the solar edge is too subtle for accurate displacement detection along this direction. Thus, we utilize the solar prominence to further correct the preliminary displacement in the horizontal direction. The structure of solar prominences is distinct and relatively stable over short periods, allowing us to employ a cross-correlation algorithm to update the preliminary displacement. Finally, we combine all the displacements obtained from the two steps to realize the well-alignment of solar prominence images. The specific operations are as follows.

*3.2.1. Calculate the Relative Translations for Preliminary Alignment*

The precondition of the OF method is that the solar features being tracked do not undergo significant evolution between the image pairs that need to be aligned, so we calculate the relative translations with the OF field frame by frame within the time series. The relative translations are the displacements of the next frame image relative to the previous frame one. To accurately determine the relative translations for each image pair, we first employ the cross-correlation algorithm to obtain the initial offsets $(x_{c1}, y_{c1})$ of the second image relative to the first one. We then correct the second image using these offsets. Although the accuracy of these initial offsets might be limited, as discussed in the Introduction, this step is crucial for reducing large movements between the images. After correction, the first frame and the adjusted second image form a new image pair for subsequent processing. Subsequently, we extract the overlapping areas of the solar surface from the image pair and calculate their movement field using the OF method. The left panel of Figure 6 illustrates the reference image and the OF vector field for the solar surface region of the image pair.

Since the sequence images are captured using the same observation channel, the primary distortions are due to translation transformation. Therefore, we just need to statistic the components of each vector on the X and Y axes, respectively, and consider the center of gravity position of these components as the relative offsets $(x_{o1}, y_{o1})$ for the image pair. The right panel of Figure 6 displays the histogram statistics of the vector field in the *X* and *Y* directions, respectively. Following this, we apply the relative offsets to transform the second image, using it as a new reference for the subsequent alignment process. In the same way, we repeat this "align with running reference image" process to calculate the relative translations for the entire time-series images, such as the displacements of cross-correlation $(x_{c2}, y_{c2})... (x_{ci}, y_{ci})$ and OF $(x_{o2}, y_{o2})... (x_{oi}, y_{oi})$, and so on. The total relative translations for each sequence image relative to its previous image are calculated as $(x_i = x_{ci} + x_{oi}, y_i = y_{ci} + y_{oi})$. The offsets of preliminary alignment relative to the first reference image are accumulated as $(dx_i = x_1 + x_2 + ... + x_i, dy_i = y_1 + y_2 + ... + y_i)$.

*3.2.2. Obtain the System Displacements for Precision Alignment*

After the preliminary alignment with the relative translations, we found the aligned sequence prominence images still have system deviations, such as the image drift along and vertical the solar edge, which is primarily due to two factors: first, cumulative errors in the frame-by-frame alignment method, and second, the solar rotation that causes long-term image drift. Although the NVST is equipped with a derotate device, slight rotation residues persist. Therefore, all the consecutive images finally must be further aligned directly with a reference frame to eliminate this system deviation. In our workflow, we designate the first frame of the time series as the reference image and calculate the offsets of the subsequent frames relative to it, which we consider as the systematic displacements.

To mitigate the issue of image drift and enhance the alignment of the sequence images, we calculate the systematic displacements for the initially aligned images separately in both the vertical and horizontal directions. In the vertical direction, we utilize the intensity gradient of the solar edge to determine its position in each image. After preprocessing, the solar edges are oriented in the "horizontal" direction, as depicted in Figure 5 (M'). We project the solar edges onto the Y-axis (i.e., summing the image along the horizontal direction) and calculate the gradient of the projection curve. The position of the minimum gradient indicates the solar edge coordinate on the Y-axis. The relative deviation of these coordinates between the reference and other sequence images represents the systematic drift displacements ($dy_i'$) in the vertical direction. To correct the horizontal systematic drift, we focus on the prominence regions of the images, as shown in the boxed area of Figure 5. The cross-correlation algorithm is used to track the translation of these structures. Initially, we extract the overlapping prominence areas from the reference and other sequence images and normalize them by their respective mean intensities. Subsequently, we employ a subpixel cross-correlation method to calculate the horizontal systematic drift displacements ($dx_i'$) for the image pair.

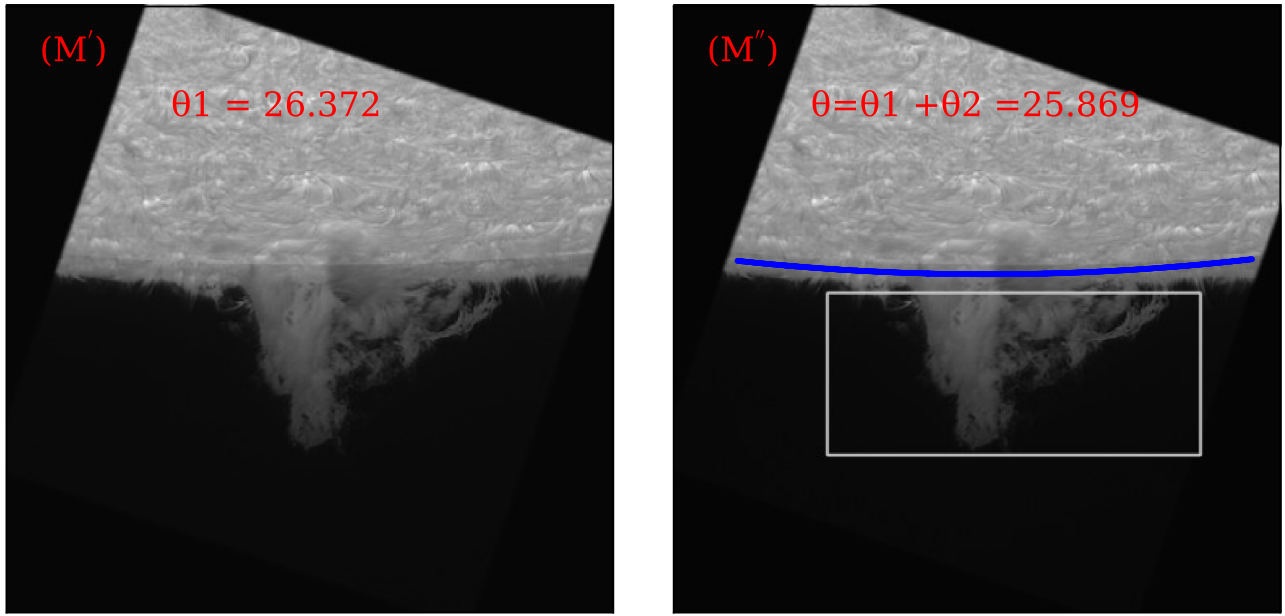

**Figure 5.** The rotation images of preprocessing. The left panel is a coarse rotation image and its angle, and the right panel is a fine rotation image and the total angle.

### 3.2.3. Create the Transform Matrices

Through our preliminary and precision alignment procedures, we determine the total offsets of translation transformation for each sequence image relative to the reference frame. These offsets are calculated as $dX_i = dx_i + dx_i'$ for the horizontal direction and $dY_i = dy_i + dy_i'$ in the vertical directions, where $dx_i$ and $dy_i$ are the offsets from the preliminary alignment, and $dx_i'$ and $dy_i'$ are the offsets from the precision alignment. However, the offsets of all sequence images should ideally show a gradual trend, which is caused by the telescope's tracking error, accompanied by a small number of random variables due to the influence of atmospheric turbulence. To minimize calculation errors and enhance alignment accuracy, we apply the RANSAC method to perform a curve fit on the offsets ($dX_i, dY_i$). The fitting values obtained from this process are considered the ultimate offsets for the translation transformation. Additionally, we combine these fitting offsets with the image rotation angle calculated during the preprocessing stage to construct a series of transformation matrices as follows:

$$\begin{pmatrix} \cos\theta & -\sin\theta & dX_i \\ \sin\theta & \cos\theta & dY_i \\ 0 & 0 & 1 \end{pmatrix}. \qquad (2)$$

By applying the corresponding transformation matrix to each sequence image, we ensure that all monochromatic sequence prominence images are precisely aligned with the reference image.

### 3.3. Alignment of Imaging Spectroscopic Prominence Images

The imaging spectroscopic observations can deliver abundant spectrum information on solar activities, playing an important role in advancing solar physics research. We have developed a high-precision alignment approach for the Hα imaging spectroscopic data acquired by the NVST, as detailed in our previous work (Y.-F. Cai et al. 2022). However, since the prominence structures of images are too weak to align the far-winged prominence images. We need to make some improvements to our previous alignment techniques for the imaging spectroscopic prominence images.

After the processing of Section 3.2, the sequence prominence images at a single wavelength (usually the images of Hα center line) have been aligned. To further align the imaging spectroscopic prominence images, we proceed with the following steps: First, we apply the aforementioned transform matrices to other off-band prominence images. The reason for this step is that the time intervals between spectroscopic prominence images within a single wavelength scan are very close, leading to similar translation transformations among them. Second, we employ an OF-based alignment method to register the winged prominence images. This method involves several steps: calculating the OF vector fields for the symmetrical off-band images relative to the Hα center line image; counting histograms for each *x*- and *y*-component of the OF vector fields and determining the corresponding proportion of effective points; and utilizing the effective point proportions as weighting coefficients in the least-squares method to derive the movement offsets for each off-band image relative to each other. Third, the solar edges are consistent across all imaging spectroscopic prominence images. Thus, we use their gradients to further refine the movement offsets. Ultimately, the imaging spectroscopic prominence images are well-aligned in both the time-series and wavelength dimensions, culminating in the creation of a 4D data cube (*x*, *y*, λ, t).

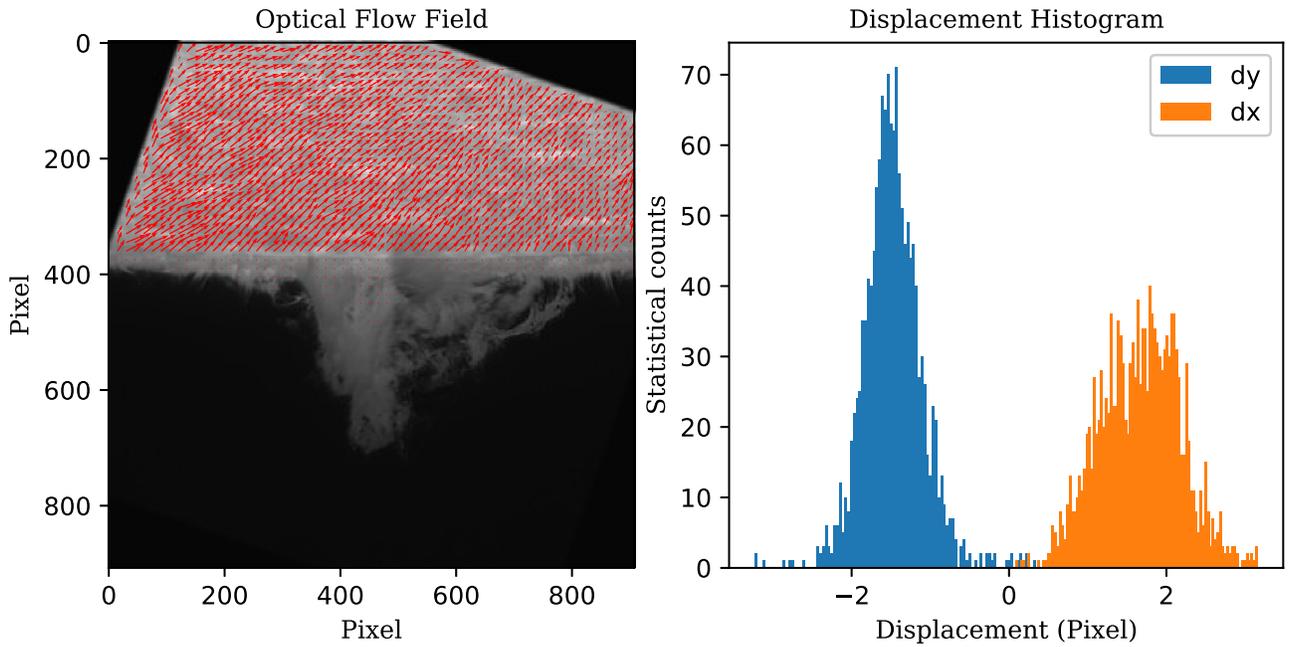

**Figure 6.** The optical flow field for the solar surface region of the image pair and their histogram statistics in *X* and *Y* directions, respectively.

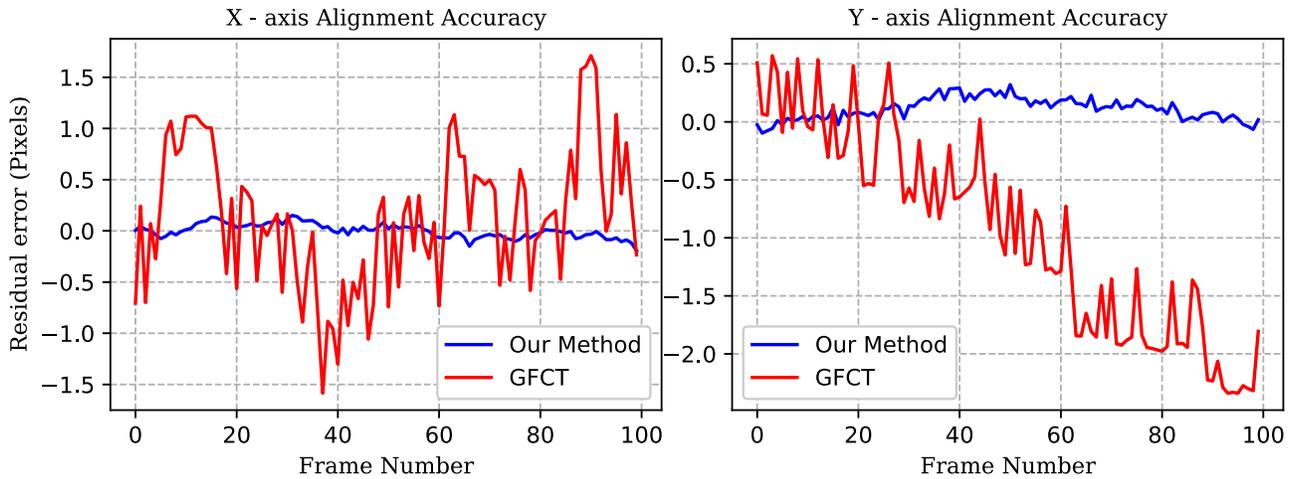

**Figure 7.** The alignment accuracy of 100 time-sequence prominence images using two different methods. The red and blue solid lines represent the performance of the GFCT and our proposed methods, respectively. The residual errors of x- and y-directions are plotted in two panels.

## 4. Alignment Accuracy and Result

To validate the alignment accuracy and result for the NVST's monochromatic time-series prominence images, we first generate 100 sets of random 2D floating-point numbers in the *x*- and *y*-directions as the customized translation displacements. These are applied to a Hα line center prominence image observed by NVST on 2016 June 4, effectively creating a simulated sequence of 100 images. We then evaluate the alignment accuracy by calculating the residual errors with our proposed method between the customized and measured displacements. For comparative analysis, we also employ the GFCT method to determine the residual errors. Figure 7 illustrates the 100 residual errors obtained using both alignment methods for the *x*- and *y*-directions. The GFCT method exhibits larger residual errors, exceeding 1 pixel, and a significant cumulative error in the *y*-axis. This indicates that images aligned with the GFCT method would show noticeable oscillation and systematic drift along the solar edge, as previously discussed. In contrast, our method yields residual errors below 0.5 pixels, demonstrating superior alignment accuracy and reduced cumulative error. To visually compare the alignment results, Figure 8 presents a local region of the reference prominence image, including the solar edge and prominence area, along with the contours of the solar edge. It is evident that there is a significant misalignment between the solar edges in the reference and corrected images using the GFCT method. In contrast, the edges perfectly overlap when aligned using our method. It is important to note that only 100 frame sequence images were used here for algorithm accuracy testing. In conventional observations, where each group typically contains tens of thousands of images (about eight frames frames/s), the GFCT method becomes less reliable due to excessive cumulative errors. In conclusion, the alignment results confirm that our algorithm's reliability and robustness are superior to that of the GFCT method.

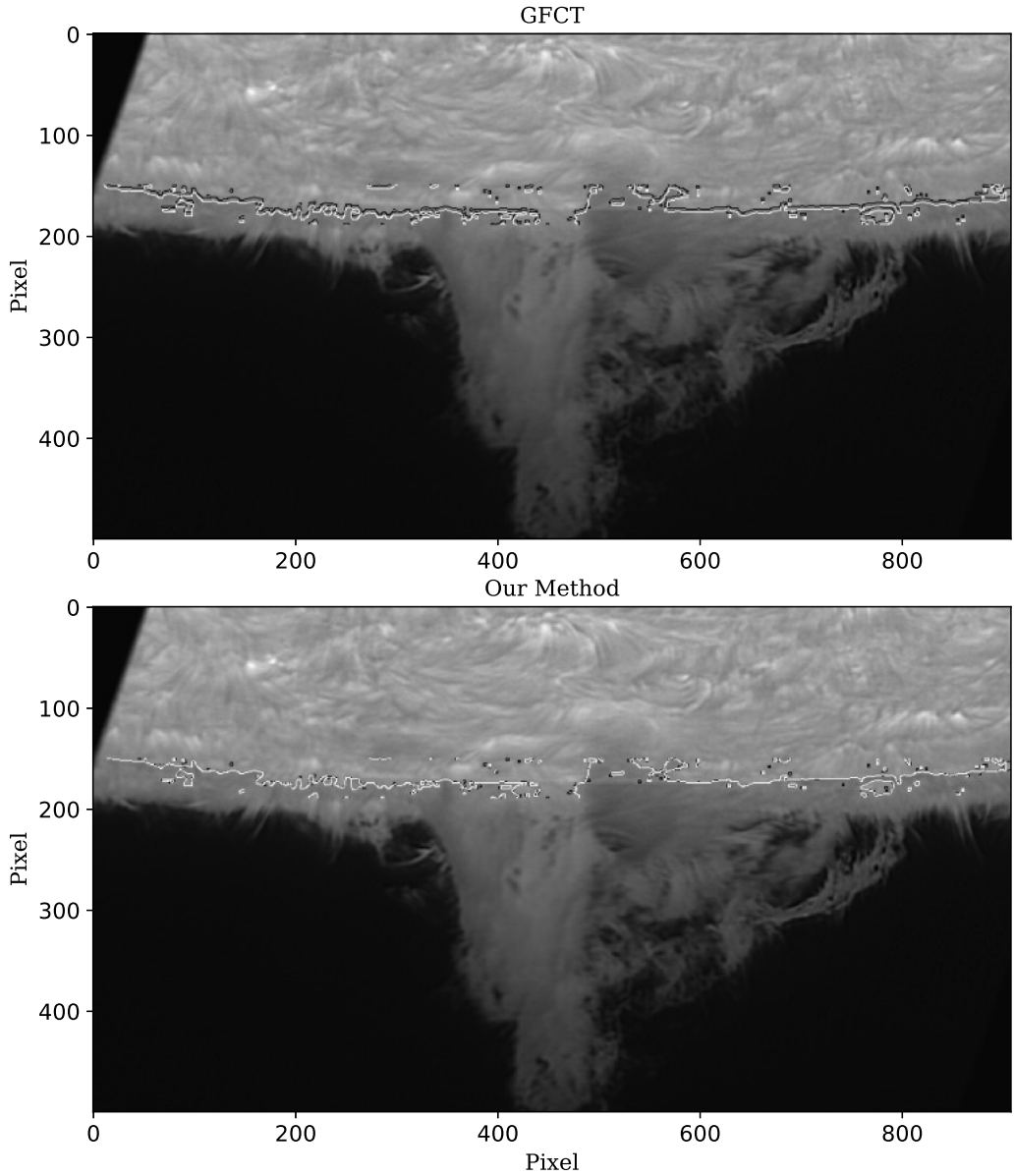

**Figure 8.** The local region of the reference prominence image and the contours of the solar edge of aligned images with two different methods. The black and white lines are the edge contours of the reference and the 100th image in the aligned sequence. The top and bottom panels are for the GFCT and our proposed methods, respectively.

For the imaging spectroscopic observations of solar prominence, we present the well-aligned images for three off-bands corresponding to the original images of Figure 2, along with their pseudo-color composite image, as shown in Figure 9. It is important to note that this data set was observed at five center wavelengths in total, and all the imaging spectroscopic images were aligned simultaneously using the method outlined in Section 3.3. Here, we display only the well-aligned images for the line center and the red off-bands (R040 and R070) to conserve the drawing space. The composite pseudo-color image, which combines these three wavelengths, effectively demonstrates the alignment results and their accuracy. The fine-scale solar structures within the solar surface and their prominence are clearly visible in the composite image. These structures would appear blurred or smoothed out if there were misalignment among the imaging spectroscopic images.

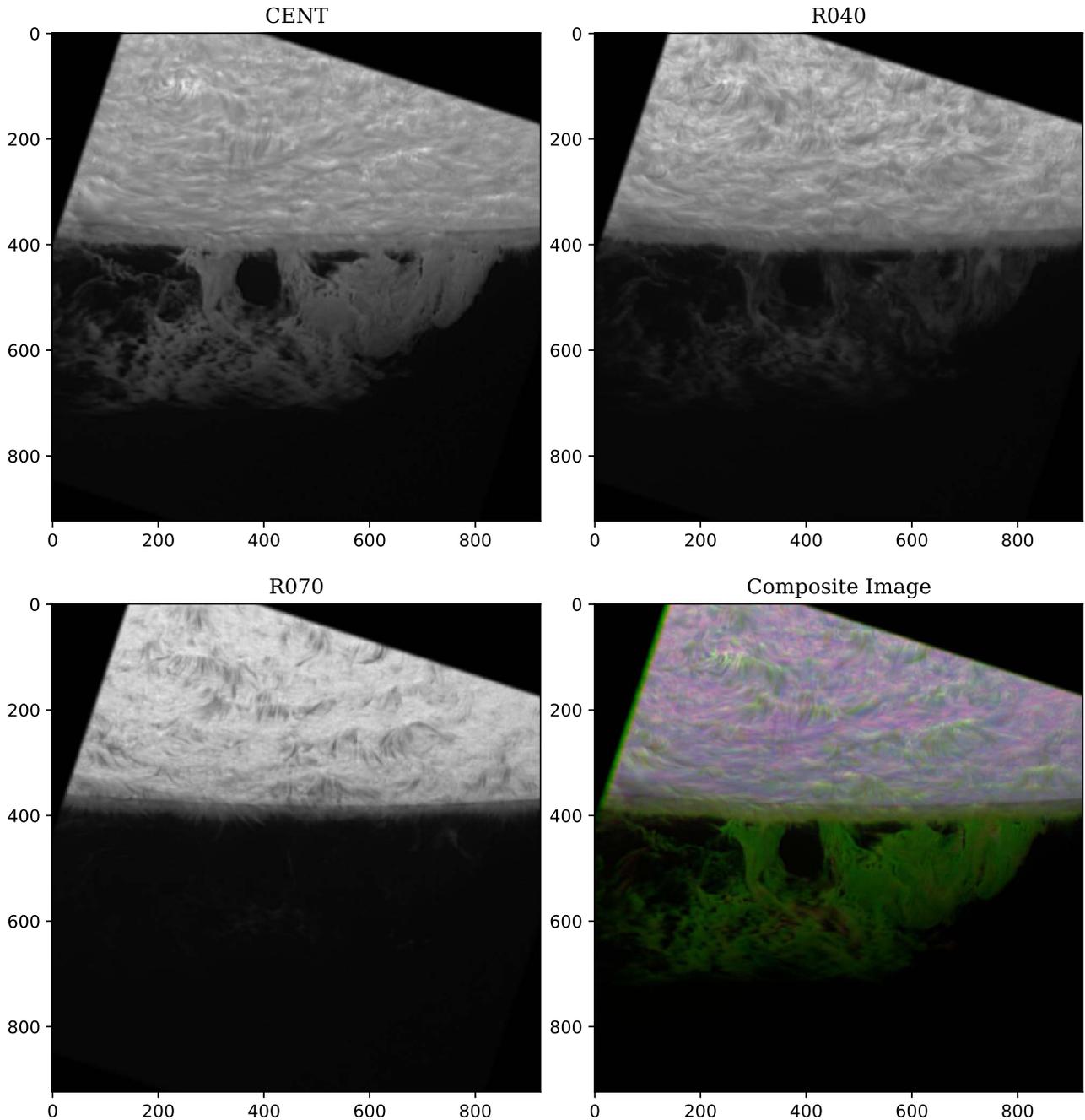

**Figure 9.** The aligned high-resolution imaging spectroscopic prominence images and their pseudo-color composite image. The three monochromatic images with different wavelengths correspond to the images in Figure 2.

## 5. Conclusion and Discussion

The NVST is equipped with a filter-based imaging system that enables high-resolution imaging and spectroscopic observations of solar prominence. However, scientists constantly encounter alignment obstacles when using our prominence data set for scientific research. In this study, we propose high-precision alignment methods for high-resolution prominence images of NVST. Based on the current observation status and characteristics of data, we decompose the prominence images into three distinct parts: the solar surface region, the solar edge, and the prominence areas. Each part is then processed with suitable algorithms to enhance the overall alignment accuracy. This strategy enables us to achieve high-precision image alignment of solar prominence, with the aligned images in a standardized posture, ultimately generating science-ready 4D data cubes ($x$, $y$, $\lambda$, t) of high-resolution prominence. Our proposed method offers unprecedented high accuracy and is notably more stable and robust than traditional alignment methods. Furthermore, the process is fully automated, eliminating the need for human intervention and significantly enhancing the alignment efficiency of high-resolution prominence images from the NVST.

Our comprehensive alignment method is suitable for the majority of NVST's high-resolution solar prominence images. However, there are certain challenging cases, such as the data depicted in image (2) of Figure 1, which features a slender prominence region with a very weak intensity, leading to complete intensity saturation in the solar surface regions. In such cases, we have to abandon the alignment step based on the OF method applied to the solar surface area. However, we can still utilize the solar edges and prominence regions with their respective methods for image alignment. Therefore, it is essential to consider the actual situation of high-resolution prominence images and select the most appropriate image structures and methods to maximize the final alignment accuracy of the prominence image.

### Acknowledgments

We appreciate all the help from the colleagues of the NVST team. This work is supported by the projects of the National Natural Science Foundation of China (NSFC) under grant Nos. 12273110, 12373115, 12073077, and 11973088; the Yunnan Province XingDian Talent Support Program; the Yunnan Top-notch Talent Training Support Program; the Yunnan Fundamental Research Projects (202401AW07006 and 202301AT070350); the "Yunnan Revitalization Talent Support Program" Innovation Team Project (202405AS350012); and the Yunnan Key Laboratory of Solar Physics and Space Science (202205AG070009).

### ORCID iDs

Yunfang Cai ● https://orcid.org/0000-0002-4956-4320
Yongyuan Xiang ● https://orcid.org/0000-0002-5261-6523
Kaifan Ji ● https://orcid.org/0000-0001-8950-3875